\documentclass[aps,prb,twocolumn,groupaddress,longbibliography,preprintnumbers]{revtex4-2} 

\usepackage{amsmath}
\usepackage{amsfonts}
\usepackage{amssymb}
\usepackage{graphicx}
\usepackage{color}
\usepackage{accents}
\usepackage[colorlinks=true, citecolor=cyan]{hyperref}
\usepackage{enumitem}

\newcommand{\Tr}{\mbox{Tr}}

\newcommand{\ket}[1]{|#1\rangle}
\newcommand{\bra}[1]{\langle #1|}

\makeatletter 
\newsavebox{\@brx}
\newcommand{\llangle}[1][]{\savebox{\@brx}{\(\m@th{#1\langle}\)}%
  \mathopen{\copy\@brx\kern-0.5\wd\@brx\usebox{\@brx}}}
\newcommand{\rrangle}[1][]{\savebox{\@brx}{\(\m@th{#1\rangle}\)}%
  \mathclose{\copy\@brx\kern-0.5\wd\@brx\usebox{\@brx}}}
\makeatother

\newlength{\dhatheight} 

\newcommand{\qed}{\nobreak \ifvmode \relax \else
      \ifdim\lastskip<1.5em \hskip-\lastskip
      \hskip1.5em plus0em minus0.5em \fi \nobreak
      \vrule height0.75em width0.5em depth0.25em\fi}

\begin{document}

\title{Perfect quantum state transfer through a chaotic spin chain via many-body scars}
\author{Shane Dooley} \email[Corresponding author: ]{dooleysh@gmail.com}
\author{Luke Johnston}
\author{Patrick Gormley}
\author{Beth Campbell}
\affiliation{School of Theoretical Physics, Dublin Institute for Advanced Studies, 10 Burlington Road, Dublin 4, Ireland.}

\date{\today}
  
\begin{abstract}
  Quantum many-body scars (QMBS) offer a mechanism for weak ergodicity breaking, enabling non-thermal dynamics to persist in a chaotic many-body system. While most studies of QMBS focus on anomalous eigenstate properties or long-lived revivals of local observables, their potential for quantum information processing remains largely unexplored. In this work, we demonstrate that \emph{perfect quantum state transfer} can be achieved in a strongly interacting, quantum chaotic spin chain by exploiting a sparse set of QMBS eigenstates embedded within an otherwise thermal spectrum. These results show that QMBS in chaotic many-body systems may be harnessed for information transport tasks typically associated with integrable models. 
\end{abstract} 
 

\maketitle
 
\section{Introduction}

Strong interactions among the constituents of a quantum many-body system typically drive thermalisation of local subsystems \cite{DAl-16,Mor-18}. This generic tendency toward thermal equilibrium renders quantum information processing in such systems extremely challenging, as quantum coherence is rapidly lost. A key challenge lies in the presence of quantum chaos -- random matrix like correlations in the energy spectrum -- which generally suppress coherent transport.

However, recent theoretical proposals \cite{Shi-17a,Tur-18a,Tur-18b} and experimental demonstrations \cite{Ber-17,Blu-21,Che-22b,Su-23a,Zha-23a,Zha-25a} have shown that certain non-thermal eigenstates, known as quantum many-body scars (QMBS), can prevent thermalisation. QMBS offer a mechanism for weak ergodicity breaking, enabling non-thermal dynamics to persist for certain special initial states, even in chaotic and strongly interacting systems \cite{Ser-21a,Mou-22a,Cha-23b}. 

A central question is whether QMBS can be leveraged to reliably process quantum information. Prior studies have primarily focused on finding new models hosting QMBS \cite{Mou-18a,Mou-18b,Sch-19,Ban-20,Bul-19,Mou-20a,McC-20,Cha-20b,Hal-21a,Log-24a,Ker-25a,Moh-25a,Kat-25a}, characterising their properties \cite{Lin-19,Lin-20,ODe-20,Bul-20,Des-22a,Lan-22a,Got-23a} or amplifying the oscillations of local observables \cite{Cho-19,Doo-20b,Doo-22a}. While these features suggest the potential for coherent information storage and manipulation, demonstrations that exploit QMBS for actual quantum information tasks are rare (exceptions include Refs. \cite{Doo-21a,Doo-23a}, in the context of quantum sensing and metrology). 
 
\begin{figure}
\includegraphics [width=0.5\textwidth]{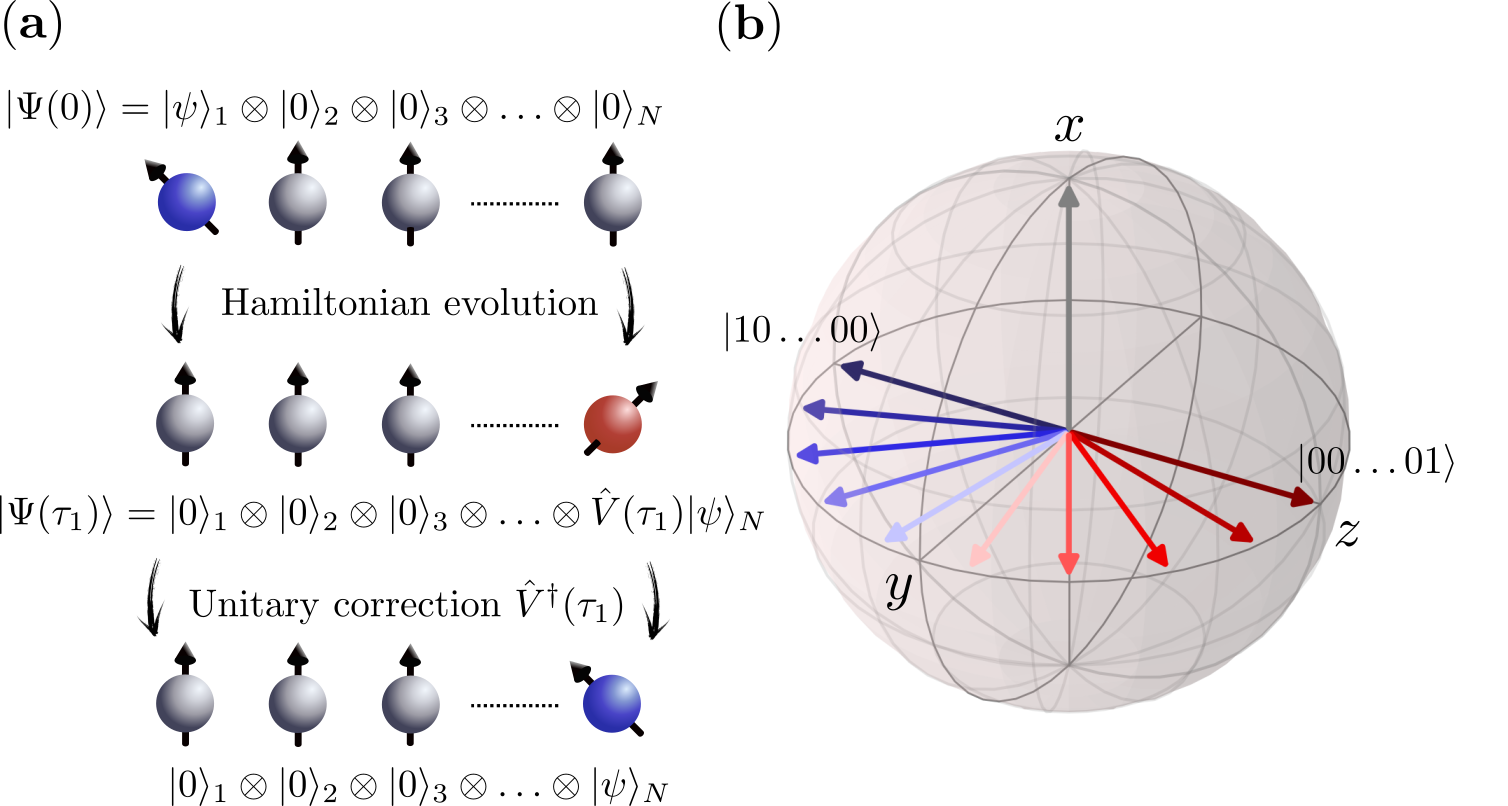} %
\caption{{\bf (a)} The aim of the quantum state transfer protocol is to transmit an arbitrary spin-1/2 state $\ket{\psi} = \alpha \ket{0} + \beta \ket{1}$ across the chain. In this paper we focus on state transfer implemented by time-independent Hamiltonian dynamics. {\bf (b)} Perfect state transfer via $\hat{H}_{\rm PST}$ can be understood intuitively as precession of an effective spin-$(N-1)/2$, embedded in the many-body Hilbert space, which transports a spin excitation across the chain. This effective spin is preserved in our many-body scarred Hamiltonian $\hat{H}_{\rm scar} = \hat{H}_{\rm PST} + \sum_n \hat{P}_n\hat{h}_n \hat{P}_n$, while the remainder of the spectrum is thermalised.}
\label{fig:schematic}
\end{figure}

For example, is quantum state transfer -- the faithful communication of a quantum state from one location to another -- possible within a quantum chaotic model? This task is not only a fundamental primitive for quantum communication and distributed quantum computing \cite{Bos-03a,Chr-04a,Chr-05a,Bos-07a,Kay-10}, but also a stringent experimental benchmark for coherent control over many-body dynamics \cite{Cha-16a,Li-18a,Xia-24a}. In this paper, we show that perfect quantum state transfer is possible in a spin chain that is both strongly interacting and quantum chaotic, yet hosts a sparse set of QMBS eigenstates. Our construction leverages these special states to implement a perfect transfer protocol, even as the bulk of the spectrum remains thermal.
 
We structure our paper as follows. In Section \ref{sec:PST} we introduce our state transfer protocol on a chain of spin-1/2 particles and show that, in the absence of noise, perfect state transfer is implemented. We then show (Section \ref{sec:chaos}) that the introduction of generic local interactions suppresses state transfer, as the chain undergoes thermalisation. In Section \ref{sec:PST_via_QMBS_spin_half} we show that by modifying the generic interaction with local projectors we can introduce QMBS to the Hamiltonian, which facilitate perfect state transfer despite the spin chain being chaotic. In essense, the subspace in which state transfer occurs is exactly spanned by the set of QMBS eigenstates, and is therefore protected against the interactions that otherwise suppress coherent transport of information. However, our demonstration of perfect state transfer through the chaotic spin-1/2 chain relies on local three-body interactions, which may be technically demanding to implement in experiments. In Section \ref{sec:PST_via_QMBS_spin1} we present another example of perfect state transfer through a spin-1 chain with only two-body couplings, which may be more typical of naturally occurring or experimentally feasible interactions.

Our results demonstrate that perfect quantum state transfer can occur in a many-body system that is quantum chaotic, challenging the conventional wisdom. In other words, by harnessing the special structure of QMBS, quantum chaos need not be an obstacle to control. 

\section{Perfect state transfer in a spin-1/2 chain}  \label{sec:PST}
 
\begin{figure*}
\includegraphics[width=\textwidth]{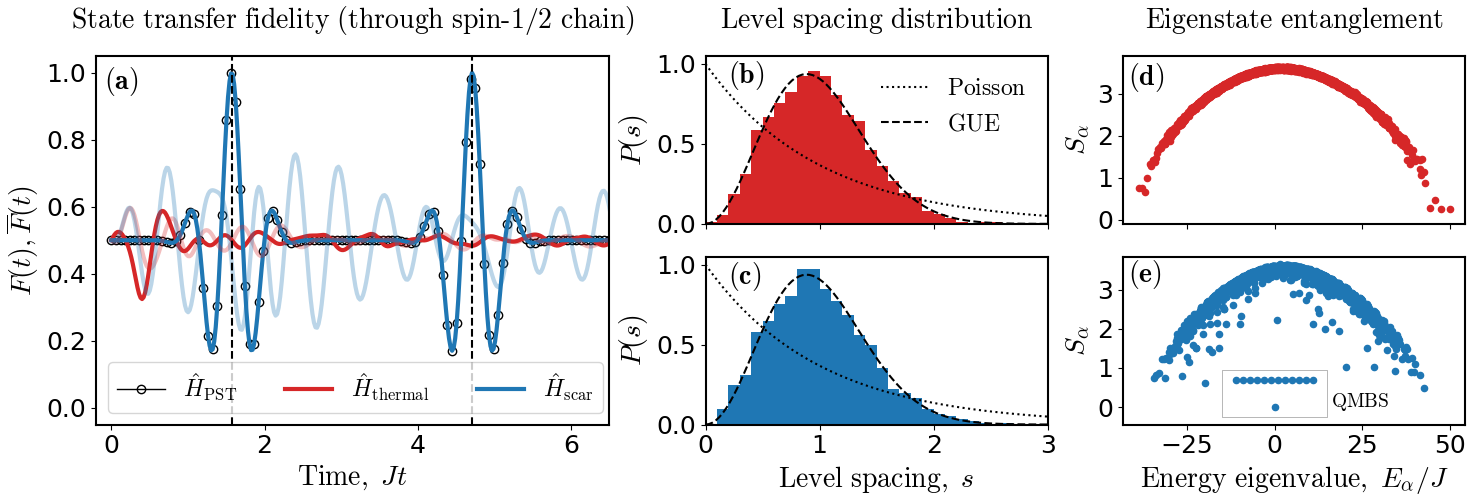}%
\caption{{\bf (a)} Time evolution of the transfer fidelity through a spin-1/2 chain, starting from either the initial state $\ket{\Psi(0)} = \ket{\psi}\ket{0}^{\otimes (N-1)}$ (heavy lines) or $\ket{\overline{\Psi}(0)} = \ket{\psi}\ket{1}^{\otimes (N-1)}$ (light lines), where $\ket{\psi} = (\ket{0}+\ket{1})/\sqrt{2}$. Evolution by the Hamiltonian $\hat{H}_{\rm PST}$ results in perfect state transfer for either initial state, at times $\tau_m = (m-1/2)\pi/\lambda$ where $m\in\{1,2,3,\hdots \}$ (vertical dashed black lines). Adding a generic local interaction $\hat{H}_{\rm thermal} = \hat{H}_{\rm PST} + \sum_n \hat{h}_n$ severely inhibits state transfer for both initial states (red lines). Adding the \emph{projected} local interaction, $\hat{H}_{\rm scar} = \hat{H}_{\rm PST} + \sum_n \hat{P}_n\hat{h}_n \hat{P}_n$ supports perfect state transfer for the initial state $\ket{\Psi(0)}$ (heavy blue line), due to the QMBS, but not for the initial state $\ket{\overline{\Psi}(0)}$ (light blue line) which is outside the QMBS subspace. {\bf (b,c)} Level spacing statistics show that both $\hat{H}_{\rm thermal}$ (b) and $\hat{H}_{\rm scar}$ (c) are chaotic. {\bf (d,e)} The entanglement entropy of the eigenstates of $\hat{H}_{\rm thermal}$ and $\hat{H}_{\rm scar}$ generally follow a volume law scaling. However, $\hat{H}_{\rm scar}$ has a small number of non-thermal eigenstates, which are responsible for the perfect state transfer despite the model being chaotic. [Other parameters: $N=12$, $\omega=0$.]} 
\label{fig:state_transfer}
\end{figure*}

We begin by introducing our state transfer protocol on a chain of $N$ spin-1/2 particle, labelled by the site index $n \in \{1, \hdots, N\}$, and with the spin-1/2 Pauli operators $\hat{X} = \ket{0}\bra{1} + \ket{1}\bra{0}$, $\hat{Y} = -i\ket{0}\bra{1} + i\ket{1}\bra{0}$ and $\hat{Z} = \ket{0}\bra{0} - \ket{1}\bra{1}$. The goal of the quantum state transfer protocol is to transmit an arbitrary spin-1/2 state $\ket{\psi} = \alpha \ket{0} + \beta\ket{1}$ from the $n=1$ spin (at initial time $t=0$) to the opposite end of the chain $n=N$ (at some later time $t=\tau_1$) -- see Fig. \ref{fig:schematic}(a) for an illustration. There are various schemes to achieve state transfer (see Refs. \cite{Bos-07a, Kay-10} for reviews). However, one of the simplest is with evolution generated by the time-independent Hamiltonian \cite{Chr-04a, Alb-04a}:
\begin{equation} \hat{H}_{\rm PST} = \frac{\omega}{2} \sum_{n=1}^N \hat{Z}_n + \frac{1}{2} \sum_{n=1}^{N-1} \lambda_n (\hat{X}_n \hat{X}_{n+1} + \hat{Y}_n \hat{Y}_{n+1}) , \label{eq:H_PST} \end{equation} where $\omega$ is a magnetic field in the $z$-direction, and the nearest-neigbour couplings are engineered to take the values $\lambda_n = \lambda\sqrt{n(N-n)}$. Starting from the initial state: \begin{eqnarray} \ket{\Psi(0)} &=& \ket{\psi} \ket{0}^{\otimes (N-1)} \nonumber \\ &=& \alpha \ket{0} \ket{0}^{\otimes (N-1)} + \beta \ket{1} \ket{0}^{\otimes (N-1)} , \label{eq:Psi_0} \end{eqnarray} Hamiltonian evolution for a time $t = \tau_1 = \frac{\pi}{2\lambda}$ leads to the state \begin{eqnarray} \ket{\Psi(\tau_1)} &=& e^{-i\tau_1\hat{H}_{\rm PST}}\ket{\Psi(0)}  \nonumber \\ &=& \ket{0}^{\otimes (N-1)} \hat{V} (\tau_1) \ket{\psi} . \label{eq:Psi_tau} \end{eqnarray} 
We see that the reduced state $\hat{V} (\tau_1) \ket{\psi}$ of the last spin-1/2 particle differs from the desired state $\ket{\psi}$ by the action of a spin-1/2 unitary $\hat{V}(\tau_1) = \exp\{i \tau_1 [\lambda(N-1) + \omega] \hat{Z}/2 + i\tau_1 \omega (N-1) \hat{\mathbb{I}}_2 \}$. However, assuming that the Hamiltonian parameters $\omega$, $\lambda$ are known, a local unitary  $\hat{V}^\dagger (\tau_1)$ can be constructed and applied to the last spin at time $\tau_1$ to correct the state. We therefore quantify the performance of the protocol with the \emph{state transfer fidelity}: \begin{equation} F(t) = \bra{\psi} \hat{V}^\dagger(t) \hat{\rho}_N(t) \hat{V}(t) \ket{\psi} , \end{equation} where $\hat{\rho}_N(t) = \Tr_{1,\hdots,N-1} \ket{\Psi (t)} \bra{\Psi(t)}$ is the reduced density matrix of the last spin in the chain. In Fig. \ref{fig:state_transfer}(a) we plot the state transfer fidelity (black markers), clearly showing $F(\tau_1) = 1$ at $t = \tau_1 = \frac{\pi}{2\lambda}$.

To intuitively understand this perfect state transfer, let us first observe that the Hamiltonian has a $U(1)$ symmetry, corresponding to the total magnetisation $\hat{M} = \sum_n \hat{Z}_n$ being a conserved quantity $[\hat{H}_{\rm PST}, \hat{M}] = 0$.
The initial state superposition in Eq. \ref{eq:Psi_0} has two components. First, the $\ket{0}^{\otimes N}$ component is an eigenstate in the subspace of maximal total magnetisation $M=N$, and therefore only accumulates a phase $\ket{0}^{\otimes N} \to e^{-itN\omega/2}\ket{0}^{\otimes N}$ during the evolution. Second, the $\ket{1}\ket{0}^{\otimes (N-1)}$ is an element of the next lowest subspace $M = N-2$, and therefore evolves entirely within this subspace, which is spanned by the states:
\begin{equation} \ket{n} \equiv \ket{0_1 \hdots 0_{n-1} 1_n 0_{n+1} \hdots 0_N} . \label{eq:n_states} \end{equation}
It is straightforward to verify that the Hamiltonian $\hat{H}_{\rm PST}$ restricted to this subspace is (up to an added term proportional to the identity in the subspace) equal to: \begin{equation} \hat{\mathsf{H}}_{\rm subspace} = 2\lambda \hat{J}^x , \label{eq:Sx} \end{equation} where $\hat{J}^+ \equiv \sum_{n=1}^{N-1} \sqrt{n(N-n)} \ket{n + 1}\bra{n}$ , $\hat{J}^x = \frac{1}{2}(\hat{J}^+ + \hat{J}^-)$, $\hat{J}^y = \frac{i}{2}(\hat{J}^- - \hat{J}^+)$ and $\hat{J}^z = \frac{1}{2}[\hat{J}^+, \hat{J}^-] = \frac{1}{2} \sum_{n=}^N (2n-N-1) \ket{n}\bra{n}$. These $\hat{J}^\mu$ are the familiar spin angular momentum operators for a large spin-$J$ particle, with $J=(N-1)/2$. Starting from the lowest weight spin state in the $z$-direction $\ket{n=1} = \ket{10...0}$, evolution by $\hat{\mathsf{H}}_{\rm subspace} = 2\lambda \hat{J}^x$ causes the effective spin-$J$ to precess around its $x$-axis to the highest weight spin state $\ket{n=N} = \ket{0...01}$ after an evolution time $t = \tau_1 = \pi/2\lambda$, implementing the transfer of the excitation across the chain [see Fig. \ref{fig:schematic}(b)]. Moreover, we expect perfect state transfer at periodic intervals thereafter, at the times $t = \tau_m = \pi(m - 1/2)/\lambda$ for $m = 1,2,3,\hdots$, as the effective spin continues to precess.

Finally for this subsection, we note that perfect state transfer can also be achieved starting from the initial state: \begin{eqnarray} \ket{\overline{\Psi}(0)} &=& \ket{\psi} \ket{1}^{\otimes (N-1)} \\ &=& \alpha \ket{0}\ket{1}^{\otimes (N-1)} + \beta \ket{1}\ket{1}^{\otimes (N-1)}.  \end{eqnarray} Following the same line of reasoning as above, the $\beta$-component $\ket{1}^{\otimes N}$ accummulates a phase, while the $\alpha$-component $\ket{0}\ket{1}^{\otimes (N-1)}$ evolves as a precessing spin-$(N-1)/2$ particle in a subspace spanned by the states $\ket{\overline{n}} = \ket{1_1 \hdots 1_{n-1} 0_n 1_{n+1} \hdots 1_N}$. For this initial state, however, the final unitary correction of the last spin should be modified $\hat{V}(\tau_1) \to \hat{V}^\dagger(\tau_1)$, and the transfer fidelity is given by $\overline{F}(t) = \bra{\psi} \hat{V}(t) \hat{\rho}_N(t) \hat{V}^\dagger(t) \ket{\psi}$.

\section{State transfer inhibited by generic local interactions} \label{sec:chaos}

There are various ways in which error may arise in the state transfer protocol. One common way is by the presence of additional unwanted terms in the Hamiltonian. To demonstrate this, let us modify the Hamiltonian to:
\begin{equation} \hat{H}_{\rm thermal} = \hat{H}_{\rm PST} + \sum_{n} \hat{h}_n , \label{eq:H_thermal} \end{equation} where $\hat{h}_n$ is a local operator with support over a few sites in the neighbourhood of $n$. Generally, the additional interaction will break the integrability of the model. For example, let us choose $\hat{h}_n = \hat{\mathbb{I}}_2^{\otimes (n-2)} \otimes \hat{h} \otimes \hat{\mathbb{I}}_2^{\otimes (N - n - 1)}$ where $\hat{h}$ is a randomly generated three-spin Hermitian operator \footnote{We generate the random three-spin Hermitian matrix as $\hat{h} = \frac{1}{2}(\hat{A} + \hat{A}^\dagger)$ where each element of the $8 \times 8$ matrix $\hat{A}$ is a complex number whose real and imaginary parts are independent Gaussian random variables with zero mean and unit variance.}. In Fig. \ref{fig:state_transfer}(b) we plot the level spacing statistics for $\hat{H}_{\rm thermal}$ corresponding to this example. The distribution closely matches that of a random matrix from the Gaussian unitary ensemble (GUE), as expected for a quantum chaotic system \cite{DAl-16}.

We also expect that adding the generic perturbation to the Hamiltonian will cause the high-energy eigenstates to thermalise. More precisely, the eigenstates of a generic chaotic Hamiltonian are expected to satisfy the eigenstate thermalisation hypothesis (ETH), meaning that all local observables take thermal expectation values in individual eigenstates \cite{Deu-91,Sre-94,Sre-99}. A direct consequence of the ETH is that local subsystems thermalise through time-evolution, even as the state of the entire system remains pure \cite{DAl-16,Mor-18}.

To show that the eigenstates $\{ \ket{E_\alpha} \}$ of $\hat{H}_{\rm thermal}$ are consistent with the ETH, in Fig.~\ref{fig:state_transfer}(d) we plot their half-chain entanglement entropy, computed via the von Neumann entropy $S_\alpha \equiv S_{\rm ent} (\hat{\rho}_\alpha) = -\Tr[\hat{\rho}_\alpha \log\hat{\rho}_\alpha]$, where $\hat{\rho}_\alpha = \Tr_{1,\hdots,\lceil N/2 \rceil} \ket{E_\alpha}\bra{E_\alpha}$ is the reduced density matrix obtained by tracing out the left half of the spin chain. For eigenstates in the middle of the spectrum, we observe high entanglement entropy, consistent with the volume-law scaling characteristic of thermal eigenstates \cite{Gar-18a}.

Given the quantum chaotic level statistics and the thermal eigenstates of $\hat{H}_{\rm thermal}$, it is unsurprising that the state transfer fidelity is significantly degraded during the evolution. This is illustrated by the red curves in Fig.~\ref{fig:state_transfer}(a), which show the fidelity of state transfer as a function of time for the two initial states $\ket{\Psi(0)}$ (heavy red line) and $\ket{\overline{\Psi}(0)}$ (light red line). In both cases, the fidelity rapidly saturates to $F \approx 1/2$, consistent with the reduced density matrix of the last spin approaching the infinite-temperature (maximally mixed) state, $\hat{\rho}_N \approx \hat{\mathbb{I}}_2 / 2$.

\section{Perfect state transfer in a chaotic spin-1/2 chain with QMBS}
\label{sec:PST_via_QMBS_spin_half}

We will now show that the introduction of QMBS to the system can facilitate perfect
state transfer despite the spin chain being chaotic. Let us consider the Hamiltonian: \begin{equation} \hat{H}_{\rm scar} = \hat{H}_{\rm PST} + \sum_{n=1}^{N-1} \hat{P}_{n} \hat{h}_n \hat{P}_n , \label{eq:H_scar} \end{equation} which differs from $\hat{H}_{\rm thermal}$ in Eq. \ref{eq:H_thermal} only by the introduction of the local projectors $\hat{P}_n = \hat{\mathbb{I}}_2^{\otimes (n-2)} \otimes \hat{P} \otimes \hat{\mathbb{I}}_2^{\otimes (N - n - 1)}$ where:
\begin{eqnarray} \hat{P} &=& \hat{\mathbb{I}}_2^{\otimes 3} - \ket{000}\bra{000} - \ket{001}\bra{001} \nonumber \\ && \qquad - \ket{010}\bra{010} - \ket{100}\bra{100} . \label{eq:P} \end{eqnarray} Just like for the Hamiltonian $\hat{H}_{\rm thermal}$ condidered previously, the second term $\sum_n \hat{P}_n\hat{h}_n\hat{P}_n$ generally breaks the integrability of the model. For example, let us again choose $\hat{h}_n = \hat{\mathbb{I}}_2^{\otimes (n-2)} \otimes \hat{h} \otimes \hat{\mathbb{I}}_2^{\otimes (N - n - 1)}$ where $\hat{h}$ is a randomly generated Hermitian operator acting on three spins. In Fig. \ref{fig:state_transfer}(c) the level spacing statistics of $\hat{H}_{\rm scar}$ for this example indicate that the Hamiltonian is quantum chaotic. At first glance, one might expect that this will prevent state transfer through the spin chain. However, the crucial difference between $\hat{H}_{\rm thermal}$ the $\hat{H}_{\rm scar}$ is that the latter Hamiltonian hosts QMBS, due to its special structure involving the local projectors $\hat{P}_n$. This is evident in the eigenstate entanglement entropies shown in Fig. \ref{fig:state_transfer}(e): the vast majority of eigenstates are thermal (they have a large entanglement) but there is a small subset of $N+1$ non-thermal eigenstates near the middle of the energy spectrum with significantly lower entanglement. These are the QMBS eigenstates. 

It is straightforward to show mathematically that the Hamiltonian $\hat{H}_{\rm scar}$ has a set of $N+1$ QMBS. First, we observe that the state $\ket{0}^{\otimes N}$, which is an eigenstate of the first term $\hat{H}_{\rm PST}\ket{0}^{\otimes N} = \frac{\omega N}{2}\ket{0}^{\otimes N}$, is annihilated by every projector $\hat{P}_n$, and therefore also by the sum $\sum_n \hat{P}_n \hat{h}_n \hat{P}_n \ket{0}^{\otimes N} = 0$. It follows that this state is a QMBS of the total Hamiltonian $\hat{H}_{\rm scar} \ket{0}^{\otimes N} = \frac{\omega N}{2}\ket{0}^{\otimes N}$. It is clearly visible as a zero-entanglement outlier in the eigenstate entanglement entropy plot in Fig. \ref{fig:state_transfer}(e).

Next, we observe that the states $\ket{n}$ (defined in Eq. \ref{eq:n_states}) are annihilated by every projector $\hat{P}_n$, i.e., $\hat{P}_n \ket{n'} = 0$ for all $n$, $n'$. The second term in Eq. \ref{eq:H_scar} therefore also annihilates these states, $\sum_{n} \hat{P}_{n} \hat{h}_n \hat{P}_n \ket{n'} = 0$, and the subspace spanned by $\{ \ket{n} \}_{n=1}^N$ is ``invisible'' to the Hamiltonian perturbation. However, we already know that the Hamiltonian $\hat{H}_{\rm PST}$ restricted to this subspace is given by $\hat{\mathsf{H}}_{\rm subspace} = 2\lambda\hat{J}^x$ (Eq. \ref{eq:Sx}). Let $\ket{x_n} = e^{i \pi \hat{J}^y / 2}\ket{n}$ be the eigenstates of $\hat{J}^x$, with the harmonically spaced eigenvalues $\hat{J}^x \ket{x_n} = \frac{1}{2}(2n - N - 1) \ket{x_n}$. It is clear that these must be eigenstates of the total Hamiltonian, with $\hat{H}_{\rm scar} \ket{x_n} = [\omega (N-2)/2 + \lambda (2n - N - 1)] \ket{x_n}$ [the $\omega (N-2)/2$ component comes from the total magnetization in the $M = N-2$ subspace]. Since the $\ket{x_n}$ are a superposition of at most $N$ product states $\ket{n}$, their half-chain entanglement entropy is bounded by the sub-volume scaling $S_{\rm ent} \leq \log\frac{N}{2}$, i.e., they are non-thermal QMBS. These are clearly visible as the band of low (but non-zero) entanglement outliers in Fig. \ref{fig:state_transfer}(e).
 
Since the initial state $\ket{\Psi(0)}$ is a superposition of the QMBS states, the dynamics relevant to perfect state transfer are entirely unaffected by the additional term $\sum_n \hat{P}_n \hat{h}_n \hat{P}_n$ in the Hamiltonian, so we expect perfect state transfer at times $t = \tau_m$ starting from this initial state. Indeed, we see this in our numerical simulation in Fig. \ref{fig:state_transfer}(a) (heavy blue line). This demonstrates that perfect state transfer is possible, despite the spin chain being quantum chaotic and the vast majority of the eigenstates being thermal. On the other hand, starting from the initial state $\ket{\overline{\Psi}(0)}$, which has no overlap with any of the QMBS eigenstates, the state transfer is highly suppressed [light blue line in Fig. \ref{fig:state_transfer}(a)].

We emphasise that the perfect state transfer generated by $\hat{H}_{\rm scar}$ does not rely on any local conserved quantities of the Hamiltonian, such as total magnetisation. In fact, the $U(1)$ symmetry present in $\hat{H}_{\rm PST}$ is broken by the added $\sum_n \hat{P}_n \hat{h}_n \hat{P}_n$ term in the Hamiltonian, and $\hat{H}_{\rm scar}$ has no obvious local conserved quantities (apart from the trivial identity operator and the Hamiltonian itself). However, the Hamiltonian does have a \emph{non-local} symmetry, $[\hat{H}_{\rm scar}, \hat{J}^x] = 0$, where $\hat{J}^x$ is the effective spin operator defined below Eq. \ref{eq:Sx}, which protects the dynamics necessary for perfect state transfer.

We note that, despite the unusual three-body structure of the projector $\hat{P}$ in Eq. \ref{eq:P}, if $\hat{h}_n$ is a one-body operator then it reduces to the simple form $\sum_n\hat{P}_n \hat{h}_n \hat{P}_n = \sum_n \ket{1}\bra{1}_{n-1} \otimes \hat{h}_n \otimes \ket{1}\bra{1}_{n+1}$. For instance, if $\hat{h}_n = \hat{X}_n$ it is the familiar PXP Hamiltonian which can be implemented in Rydberg atoms arrays \cite{Les-11,Ber-17}. However, implementing the $\hat{H}_{\rm PST}$ term experimentally alongside the PXP term may be challenging, as it couples symmetry sectors of the PXP model that are usually separated by a large energy gap due to the Rydberg blockade.

Finally, we note that for the example simulated numerically in Fig. \ref{fig:state_transfer}, the terms $\hat{P}_n \hat{h}_n \hat{P}_n$ were chosen to be homogeneous across the chain. However, this uniformity is not essential for perfect state transfer -- these terms could vary from site to site without compomising the protocol. Moreover, while our Hamiltonians in Eqs. \ref{eq:H_PST}, \ref{eq:H_thermal}, \ref{eq:H_scar} are defined for one-dimensional spin chains, the construction can be readily extended to higher-dimensional lattices \cite{Xia-24a}. 

\section{Perfect state transfer in a chaotic spin-1 chain with QMBS}
\label{sec:PST_via_QMBS_spin1}
 
\begin{figure*}
\includegraphics[width=\textwidth]{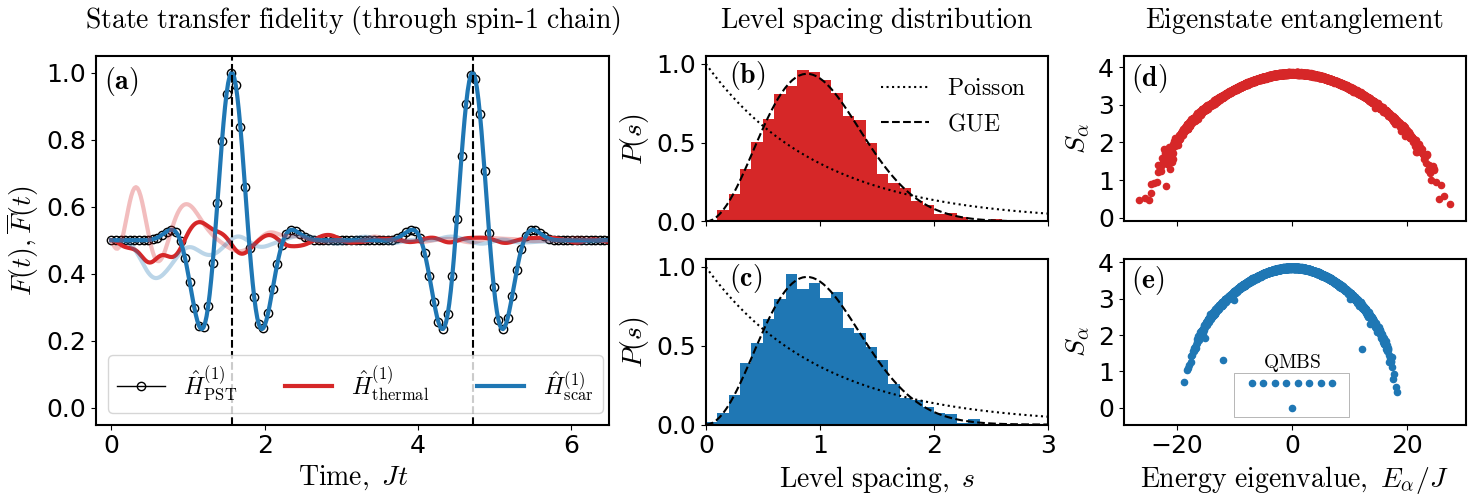}%
\caption{{\bf (a)} Time evolution of the transfer fidelity through a spin-1 chain, starting from either the initial state $\ket{\Psi(0)} = \ket{\psi}\ket{+}^{\otimes (N-1)}$ (heavy lines) or $\ket{\overline{\Psi}(0)} = \ket{\psi}\ket{-}^{\otimes (N-1)}$ (light lines), where $\ket{\psi} = (\ket{+}+\ket{-})/\sqrt{2}$. Evolution by the Hamiltonian $\hat{H}^{(1)}_{\rm PST}$ results in perfect state transfer for either initial state. Adding generic two-body local interactions (Eq. \ref{eq:H_thermal_spin1}) severely inhibits state transfer for both initial states (red lines). However, adding the \emph{projected} local interaction (Eq. \ref{eq:H_scar_spin1}) supports perfect state transfer for the initial state $\ket{\Psi(0)}$ (heavy blue line). {\bf (b,c)} Level spacing statistics show that both $\hat{H}_{\rm thermal}^{(1)}$ (b) and $\hat{H}_{\rm scar}^{(1)}$ (c) are chaotic. {\bf (d,e)} The entanglement entropy of the eigenstates of $\hat{H}_{\rm thermal}^{(1)}$ and $\hat{H}_{\rm scar}^{(1)}$ generally follow a volume law scaling. However, $\hat{H}_{\rm scar}^{(1)}$ has a small number of non-thermal eigenstates, which are responsible for the perfect state transfer despite the model being chaotic and strongly interacting. [Other parameters: $N=8$, $\omega=0$, $\hat{h}_{n,n'} = \frac{1}{2}(\hat{A}_{n,n'} + \hat{A}_{n,n'})/|n-n'|^3$ where $\hat{A}_{n,n'}$ is a two-spin operator whose matrix elements are complex Gaussian random variables with zero mean and unit variance.]} 
\label{fig:state_transfer_spin1} 
\end{figure*}

In the previous section, we demonstrated perfect state transfer in a quantum chaotic chain of spin-1/2 particles. This relied on three-body interactions, implemented via the projector $\hat{P}_n$ in Eq. \ref{eq:P}. While three-body interactions can be engineered experimentally, such as in the PXP model, they are typically challenging to realize. In this section, we show that perfect state transfer can also be achieved in a spin-1 chain governed solely by two-body interactions, which may be more natural or experimentally accessible in many physical platforms.

Let $\{\ket{-}, \ket{0}, \ket{+} \}$ be a local basis for each spin-1 particle, and define the operators $\hat{X}^{(1)} = \ket{-}\bra{+} + \ket{+}\bra{-}$, $\hat{Y}^{(1)} = -i\ket{-}\bra{+} + i\ket{+}\bra{-}$ and $\hat{Z}^{(1)} = \ket{+}\bra{+} - \ket{-}\bra{-}$, where the superscript is a reminder that, although they appear as spin-1/2 operators, they act on the local Hilbert space of a spin-1 particle. Consider the Hamiltonian:
\begin{equation} \hat{H}_{\rm PST}^{(1)} = \frac{\omega}{2}\sum_{n}\hat{Z}_n^{(1)} + \frac{1}{2}\sum_{n} \lambda_n (\hat{X}^{(1)}_n\hat{X}^{(1)}_{n+1} + \hat{Y}^{(1)}_n\hat{Y}^{(1)}_{n+1}) , \end{equation} with $\lambda_n = \lambda\sqrt{n(N-n)}$, which ``embeds'' the spin-1/2 chain Hamiltonian of Eq. \ref{eq:H_PST} into a chain of spin-1 particles. The goal of the state transfer protocol here is to transmit the state $\ket{\psi} = \alpha\ket{+} + \beta\ket{-}$ across the chain. Starting from either of the initial states:
\begin{equation} \ket{\Psi(0)} = \ket{\psi} \ket{+}^{\otimes N} , \quad \ket{\overline{\Psi}(0)} = \ket{\psi} \ket{-}^{\otimes N} , \end{equation} evolution by $\hat{H}_{\rm PST}^{(1)}$ leads to perfect state transfer at times $t = \tau_m = (m - 1/2)\pi/\lambda$, exactly analogously to the protocol for the spin-1/2 chain. Let us introduce interactions through the Hamiltonian: 
\begin{equation} \hat{H}_{\rm thermal}^{(1)} = \hat{H}_{\rm PST}^{(1)} + \sum_{n,n'} \hat{h}_{n,n'} , \label{eq:H_thermal_spin1} \end{equation} where $\hat{h}_{n,n'}$ is an arbitrary two-body Hermitian operator that acts non-trivially only on the spins at sites $n$ and $n'$. This form includes, for example, long-range interactions that decay with the spatial separation between spins $n$ and $n'$. It generally breaks the integrability of the Hamiltonian, induces eigenstate thermalisation, and suppresses coherent information transport, as demonstrated numerically for an example in Figs. \ref{fig:state_transfer_spin1}(a,b,d).

Let us also consider the Hamiltonian:
\begin{equation} \hat{H}_{\rm scar}^{(1)} = \hat{H}_{\rm PST}^{(1)} + \sum_{n,n'}\hat{P}_{n,n'}\hat{h}_{n,n'}\hat{P}_{n,n'} , \label{eq:H_scar_spin1} \end{equation} where $\hat{P}_{n,n'}$ is a local two-body projector that acts as: \begin{equation} \hat{P} = \hat{\mathbb{I}}_3^{\otimes 2} - \ket{+,+}\bra{+,+} - \ket{-,+}\bra{-,+} - \ket{+,-}\bra{+,-} , \label{eq:P_spin1} \end{equation} on the spin-1 particles at sites $n$ and $n'$ and as the identity operator on all other spins. The Hamiltonian $\hat{H}_{\rm scar}^{(1)}$ is generally chaotic, as shown in Fig. \ref{fig:state_transfer_spin1}(c). However, it also hosts a set of $N+1$ QMBS, visible as eigenstates with low entanglement entropy in Fig. \ref{fig:state_transfer_spin1}(e). To understand the QMBS, we observe that the state $\ket{+}^{\otimes N}$ as well as the states: \begin{equation}  \hat{X}_n^{(1)} \ket{+}^{\otimes N} = \ket{+}^{\otimes (n-1)} \ket{-} \ket{+}^{\otimes (N-n-1)} ,  \end{equation} for $n \in \{ 1, \hdots, N\}$, are all annihated by the projector $\hat{P}_{m,m'}$ for any $m$, $m'$, and are therefore also annihilated by the interaction term $\sum_{n,n'}\hat{P}_{n,n'}\hat{h}_{n,n'}\hat{P}_{n,n'}$ in the Hamiltonian. The subspace spanned by these states is therefore invisible to the interaction term, and only sees the first term $\hat{H}_{\rm PST}^{(1)}$. The initial state $ \ket{\Psi(0)} = \ket{\psi} \ket{+}^{\otimes N}$ is an element of the QMBS subspace, while the initial state $\ket{\overline{\Psi}(0)} = \ket{\psi} \ket{-}^{\otimes N}$ is outside the QMBS. We therefore expect perfect state transfer starting from $\ket{\Psi(0)}$, but not from $\ket{\overline{\Psi}(0)}$, as is demonstrated numerically for our example in Fig. \ref{fig:state_transfer_spin1}(a) (heavy blue line vs. light blue line).

Finally, we observe that for spin-1 particles, any two-body interaction $\hat{h}$ can be decomposed as a linear combination of $3^2 \times 3^2 = 81$ linearly independent Hermitian operators (i.e, 81 different interaction types). The effect of the projectors in $\hat{P}_{n,n'}\hat{h}_{n,n'}\hat{P}_{n,n'}$, with $\hat{P}_{n,n'}$ given by Eq. \ref{eq:P_spin1}, is to annihilate 45 of the possible interaction types, while commuting with the remaining 36 interaction types. So, our state transfer protocol is robust to 36 of the possible 81 kinds of two-spin interactions. An important direction of future work might be to combine this partial robustness with other error mitigation techniques \cite{Ash-15a}. For example, using dynamical decoupling, it may be possible to drive the system away from the interaction types against which the protocol is fragile and towards the interaction types against which the protocol is robust. The pulse design requirements are likely to be less severe than, for example, if all interaction types were required to be suppressed \cite{Pog-24a}.  

\section{Discussion}

We have demonstrated that perfect quantum state transfer is possible through a spin chain even in the presence of quantum chaos, in a particular set of Hamiltonians hosting QMBS. The result is striking, as chaotic dynamics are typically associated with rapid delocalisation and information scrambling, which would intuitively seem to hinder coherent state transfer. The presence of QMBS enables non-ergodic dynamics from a small subset of initial conditions, within an otherwise chaotic system, thereby supporting perfect state transfer.

Our findings point to a novel and potentially practical application of quantum many-body scars—a field that has been largely focused on characterising their structure and dynamics, with few proposals for functional use. This work suggests that QMBS may be harnessed to engineer robust quantum protocols in regimes where conventional approaches fail.

Our scarred Hamiltonians in Eqs. \ref{eq:H_scar} and \ref{eq:H_scar_spin1} bear a strong resemblance to the class of scarred Hamiltonians introduced by Shiraishi and Mori \cite{Shi-17a}. Despite these similarities, in Appendix \ref{app:relation_to_SM} we show that our Hamiltonians do not fall exactly into this class. The Shiraishi-Mori construction \emph{could} be used to embed the integrable spin-1/2 state transfer Hamiltonian in a larger chaotic spin-1 chain in a relatively trivial manner that differs from our approach. However, in Appendix \ref{app:trivial_embedding} we show that this would give an unnecessarily large scar subspace, while our scarred Hamiltonian is ``minimal'': it has exactly the scar subspace required for perfect state transfer, without any superfluous scar degrees of freedom, which allows us to find a larger set of interactions to which the state transfer protocol is robust.

We note that our scarred models feature an embedded subspace with an $SU(2)$ structure that protects the coherent dynamics underlying the state transfer. Several previously known scarred models, including the PXP model, also exhibit an exact or approximate $SU(2)$ structure \cite{Cho-19, ODe-20}. It would be interesting to explore whether other $SU(2)$-based QMBS systems could similarly support quantum information tasks such as state transfer, perhaps through solitonic excitations \cite{Ker-25a-arxiv}. Also, while QMBS can protect against certain perturbations, they may themselves be fragile under more generic interactions (see Appendix \ref{app:perturbations}). Enhancing their robustness, perhaps through dynamical decoupling or other control techniques, remains an important open challenge.

\begin{acknowledgments}
This publication has emanated from research conducted with the financial support of Taighde \'{E}ireann – Research Ireland under Grant number 22/PATH-S/10812. The data that support the findings of this article are openly available at \cite{Doo-25d}.
\end{acknowledgments}

\bibliography{/Users/dooleysh/Google_Drive/physics/papers/bibtex_library/refs}

\appendix  
\section{Relationship of our models to the Shiraishi-Mori projector embedding construction} \label{app:relation_to_SM}

The construction of our scarred Hamiltonians $\hat{H}_{\rm scar}$ (Eq. \ref{eq:H_scar}) and $\hat{H}_{\rm scar}^{(1)}$ (Eq. \ref{eq:H_scar_spin1}) is similar to, and inspired by, the projector-embedding method, due to Shiraishi and Mori \cite{Shi-17a}. However, our model does not fit exactly into their class of models.

The Shiraishi-Mori method builds a local Hamiltonian of the form $\hat{H} = \hat{H}' + \sum_n \hat{P}_n \hat{h}_n \hat{P}_n$ where $[\hat{H}', \hat{P}_n] = 0$ for all $n$. The scar subspace is the set of states $\mathcal{S} = \{ \ket{\Psi} : \hat{P}_n\ket{\Psi} = 0, \forall n \}$ that are simultaneously annihilated by all local projectors $\hat{P}_n$. Our model has the similar form $\hat{H} = \hat{H}_{\rm PST} + \sum_n \hat{P}_n \hat{h}_n \hat{P}_n$, but without the commutation relation being satisfied, i.e., $[\hat{H}_{\rm PST}, \hat{P}_n] \neq 0$. Instead, our projectors commute with the non-local operator $\hat{\mathsf{H}}_{\rm subspace} = 2 \lambda \hat{J}^x$, which is the Hamiltonian $\hat{H}_{\rm PST}$ restricted to its subspace of total magnetization $M=N-2$ (i.e., we have $[\hat{\mathsf{H}}_{\rm subspace}, \hat{P}_n] = 0$ for all $n$). Also, our scar subspace is not $\mathcal{S}$, rather it is the subspace of $\mathcal{S}$ in the $M=N-2$ and $M=N$ magnetisation sectors. So, although our construction has similarities to the Shiraish-Mori projector embedding method, there are also several differences.

\section{``Trivial'' embedding of spin-1/2 state transfer Hamiltonian in a spin-1 chain} \label{app:trivial_embedding}

It would be possible to embed the integrable spin-1/2 state transfer Hamiltonian $\hat{H}_{\rm PST}$ in a larger chaotic spin-1 chain in a relatively trivial manner by the Shiraishi-Mori projector embedding method. In our spin-1 example of Sec. \ref{sec:PST_via_QMBS_spin1}, this would be achieved by replacing the projector: \begin{equation} \hat{P} = \hat{\mathbb{I}} \otimes \hat{\mathbb{I}} - \ket{+,+}\bra{+,+} - \ket{-,+}\bra{-,+} - \ket{+,-}\bra{+,-} , \nonumber \end{equation} (given in our Eq. \ref{eq:P_spin1}) with the projector: \begin{eqnarray} \hat{P}^{(\rm triv)} &=& \hat{\mathbb{I}} \otimes \hat{\mathbb{I}} - \ket{+,+}\bra{+,+} - \ket{-,+}\bra{-,+} \nonumber \\ && \qquad - \ket{+,-}\bra{+,-} - \ket{-,-}\bra{-,-} . \nonumber \end{eqnarray} However, this replacement would give a scar subspace $\mathcal{S} = {\rm span} \{ \ket{i_1 i_2 \hdots i_N } : i_j \in \pm \}$, which is $2^N$-dimensional, when only an $(N+1)$-dimensional subspace is required to implement perfect state transfer. In this sense, the ``trivial'' construction would provide an unnecessarily large scar subspace, while our embedding is ``minimal'': it gives exactly the required scar subspace for perfect state transfer, without any superfluous scar degrees of freedom. Moreover, the trivial embedding method would not work at all in our spin-1/2 chain example in Sec. \ref{sec:PST_via_QMBS_spin_half}, since then the $2^N$-dimensional scar subspace would be the entire $2^N$-dimensional Hilbert space of the chain. An advantage of our ``minimal'' scar subspace embedding is that allows us to identify a larger set of local interaction terms ($\hat{P}_n \hat{h}_n \hat{P}_n$ in our notation) to which our protocol is robust, compared to the trivial embedding. For example, state transfer through our spin-1 chain, using the local projectors $\hat{P}$ is explicitly robust to local interactions of the form: \begin{eqnarray} \hat{h} &=& a \ket{-,-} \bra{0,+} + b \ket{-,-} \bra{+,0} + c \ket{-,-} \bra{0,-} \nonumber \\ && \qquad + d \ket{-,-} \bra{-,0} + f \ket{-,-} \bra{0,0} + {\rm h.c.} , \nonumber \end{eqnarray} since for this form of $\hat{h}$ we have $\hat{P}\hat{h}\hat{P} = \hat{h}$. However, $\hat{P}^{(\rm triv)}\hat{h}\hat{P}^{(\rm triv)} \neq \hat{h}$ for the corresponding local projectors $\hat{P}^{(\rm triv)}$ of the trivial embedding, making the robustness of state transfer to interactions of type $\hat{h}$ difficult to see for the trivial embedding.

\section{State transfer fidelity in the presence of non-projected perturbations} \label{app:perturbations}

As discussed at the end of Sec. \ref{sec:PST_via_QMBS_spin1}, perfect state transfer through the spin-1 chain is robust to 36 of the possible 81 types of two-spin interaction. In this section we examine the transfer fidelity as a function of perturbation strength, for perturbations that can destroy the QMBS.

\begin{figure*}[h]
  \centering
  \includegraphics[width=\columnwidth]{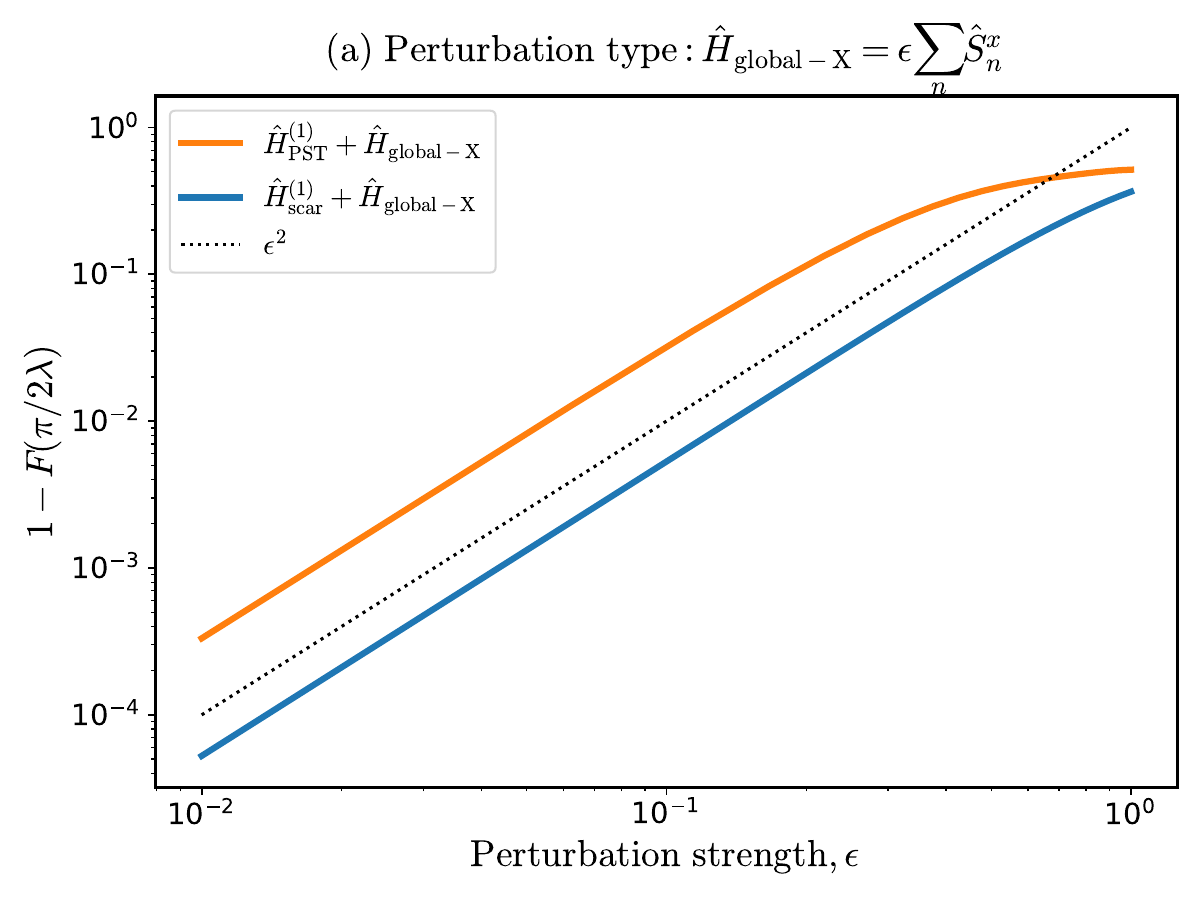}
   \includegraphics[width=\columnwidth]{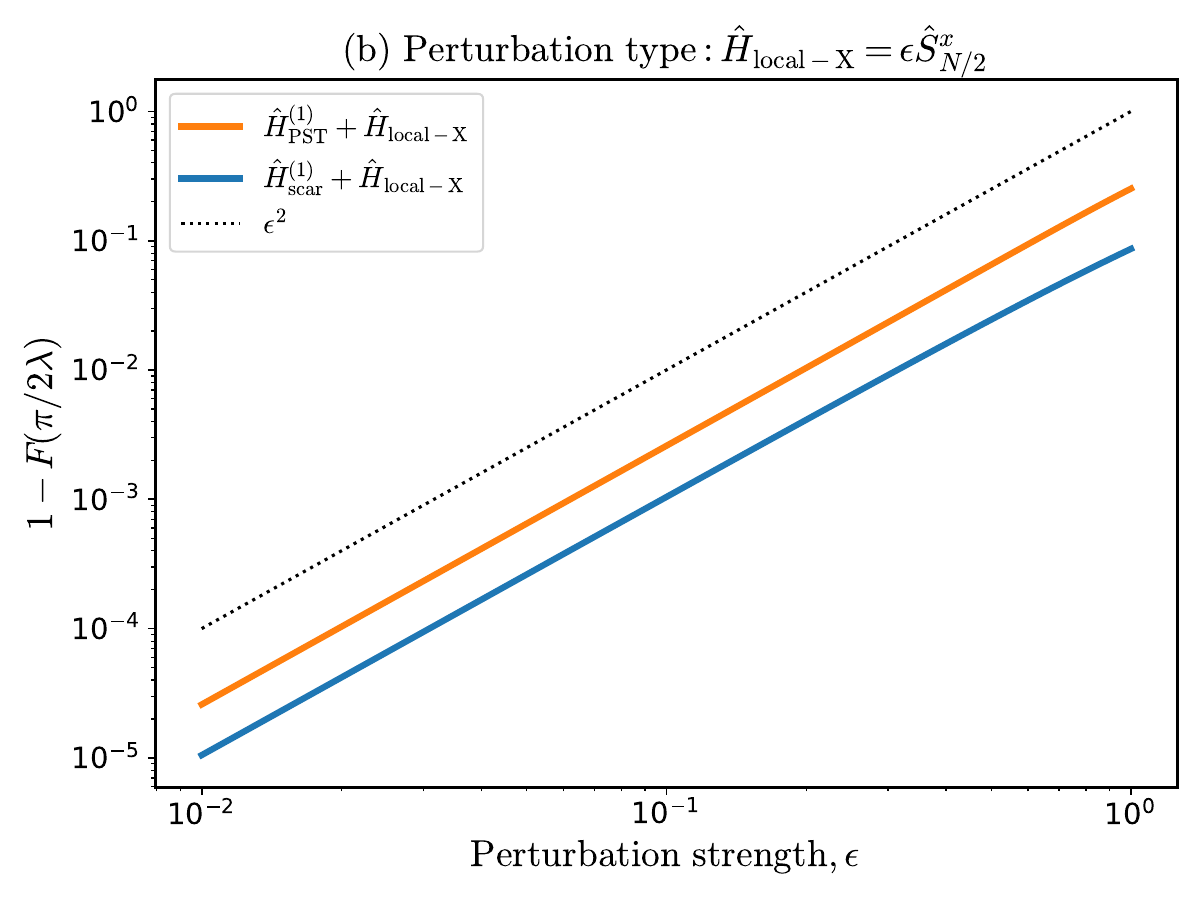}
   \includegraphics[width=\columnwidth]{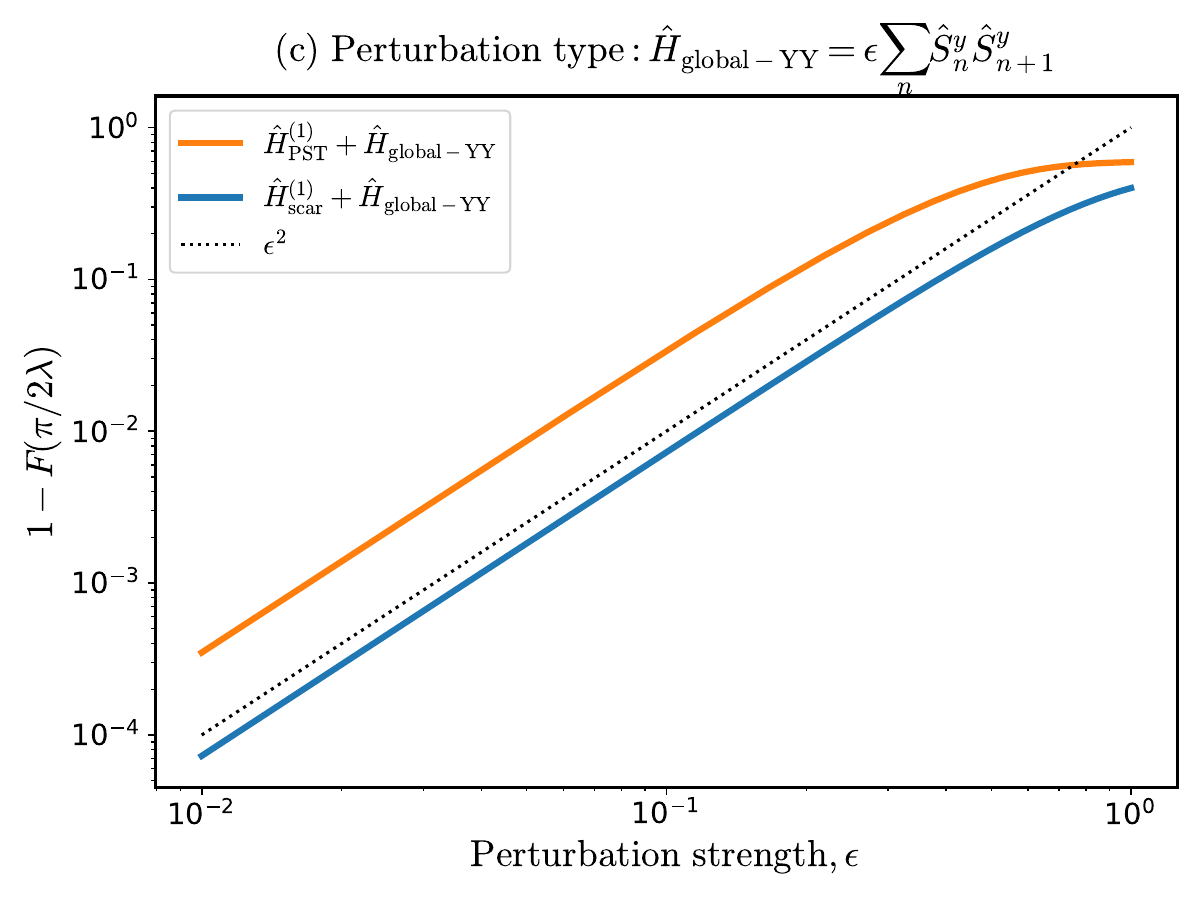}
  \caption{The state transfer infidelity, at time $t = \pi/2\lambda$, in the presence of three different perturbation types: (a) $\hat{H}_{\rm global-X} = \epsilon\sum_n \hat{S}_n^x$, (b) $\hat{H}_{\rm local-X} = \epsilon \hat{S}_{N/2}^x$, (c) $\hat{H}_{\rm global-YY} = \epsilon\sum_n \hat{S}_n^y \hat{S}_{n+1}^y$. Here, $N=8$ and all other initial conditions and parameters in $\hat{H}_{\rm scar}^{(1)}$ are generated in the same way as in Fig. \ref{fig:state_transfer_spin1}.}
  \label{fig:infidelity}
\end{figure*}

We consider our spin-1 chain model $\hat{H}_{\rm scar}^{(1)}$ (given in Eq. \ref{eq:H_scar_spin1} of our manuscript) and three different perturbations: $\hat{H}_{\rm local-X} = \epsilon \hat{S}^x_{N/2}$, $\hat{H}_{\rm global-X} = \epsilon \sum_n \hat{S}^x_{n}$, and $\hat{H}_{\rm global-YY} = \epsilon \sum_n \hat{S}^y_{n}\hat{S}^y_{n+1}$ where: \begin{equation} \hat{S}^x = \left(\begin{array}{ccc} 0 & 1 & 0 \\ 1 & 0 & 1 \\ 0 & 1 & 0 \end{array}\right) , \quad \hat{S}^y = \left(\begin{array}{ccc} 0 & -i & 0 \\ i & 0 & -i \\ 0 & i & 0 \end{array}\right) , \end{equation} in the spin-1 basis $\{\ket{-},\ket{0},\ket{+} \}$. For each perturbation type, we numerically compute the state transfer infidelity $1-F(t)$ at the time $t = \tau_1=\pi/2\lambda$ (which is the  time of perfect state transfer when $\epsilon = 0$). The blue lines in Figs. \ref{fig:infidelity}(a,b,c) show that the infidelity scales as $1 - F(\tau_1) \sim \epsilon^2$ with the perturbation strength $\epsilon$, for all perturbation types. For comparison, we also plot the infidelity as a function of $\epsilon$ when the perturbation is added to the integrable perfect state transfer Hamiltonian $\hat{H}_{\rm PST}^{(1)}$ (given in Eq. 12 of our manuscript). The orange lines in Figs. \ref{fig:infidelity}(a,b,c) show the scaling $1 - F(\tau_1) \sim \epsilon^2$ for all perturbation types in this case too. Interestingly, the state transfer fidelity is worse when the perturbation is added to the integrable Hamiltonian than when it is added to the non-integrable scarred Hamiltonian. The reason for this is unclear and deserves further study.

\end{document}